\documentclass[a4paper, oneside, twocolumn, notitlepage, 10pt]{extarticle_ecoc}
\usepackage{ecoc}
\usepackage{IEEEtrantools}
\usepackage{caption}
\usepackage{subcaption}

\begin{document}
\selectlanguage{english}

\title{Learned Digital Back-Propagation for Dual-Polarization Dispersion Managed Systems}%

\author{
    Mohannad Abu-romoh\textsuperscript{(1)}, Nelson Costa\textsuperscript{(2)},
    Antonio Napoli\textsuperscript{(3)}, Bernhard Spinnler\textsuperscript{(3)}, \\ Yves Jaou\"en\textsuperscript{(1)}, Mansoor Yousefi\textsuperscript{(1)}
}

\maketitle                  % Create title and author

%------------------------------------------ Description of Authors ----------------------------------------------%

\begin{strip}
 \begin{author_descr}

\textsuperscript{(1)} T\'el\'ecom Paris, Palaiseau, France, \textcolor{blue}{\uline{mohannad.aburomoh@telecom-paris.fr}} 

\textsuperscript{(2)} Infinera, Unipessoal Lda, Carnaxide, Portugal ~~ \textsuperscript{(3)} Infinera, Munich, Germany

 \end{author_descr}
\end{strip}

\setstretch{1.1}
%-------------------------------------------------- Footnote -------------------------------------------------------%
\renewcommand\footnotemark{}
\renewcommand\footnoterule{}
%\let\thefootnote\relax\footnotetext{text}

%-------------------------------------------------- Abstract ---------------------------------------------------------%

\begin{strip}
  \begin{ecoc_abstract}
    Digital back-propagation (DBP) and learned DBP (LDBP) are proposed for nonlinearity mitigation in WDM dual-polarization dispersion-managed systems. LDBP achieves $Q$-factor improvement of 1.8 dB and 1.2 dB, respectively, over linear equalization and a variant of DBP adapted to DM systems. 
    
%\textcopyright2022 The Author(s)
  \end{ecoc_abstract}
\end{strip}

%-------------------------------------------------- Introduction Section -------------------------------------------------------%

\section{Introduction}
Digital signal processing (DSP) can be applied in combination with coherent detection to equalize channel distortions in optical communication systems.
Examples include\cite{DSP_savory} digital back-propagation (DBP) \cite{Ezra_Ip_DBP,Antonio_DBP}, Volterra series equalization, and learned DBP (LDBP) using neural networks (NNs)\cite{Hager_butler_LDBP,Opt_express_LDBP,Hager_PBDL,Hagers_first,APTL_NN,Hager_PBDL}.
In LDBP, a NN equalizer is considered based on the computational graph generated by the split-step Fourier method
(SSFM), and optimized using the stochastic gradient descent (SGD). 
The aforementioned equalization methods were mostly developed for chromatic dispersion uncompensated links. 

However, before the advent of the coherent detection, dispersion needed to be compensated in the optical domain. These links correspond to dispersion-managed (DM) systems where dispersion compensating fibers (DCF) are deployed in intermediate nodes, such as optical amplification sites, to compensate for the accumulated chromatic dispersion. These DM systems are still widely used since the removal of the DCFs may not be cost effective or even impossible for various reasons (e.g., in submarine cables or when legacy on-off keying (OOK) signals are still used). 

\begin{figure*}[htp]
    \centering
    \includegraphics[width=1\linewidth]{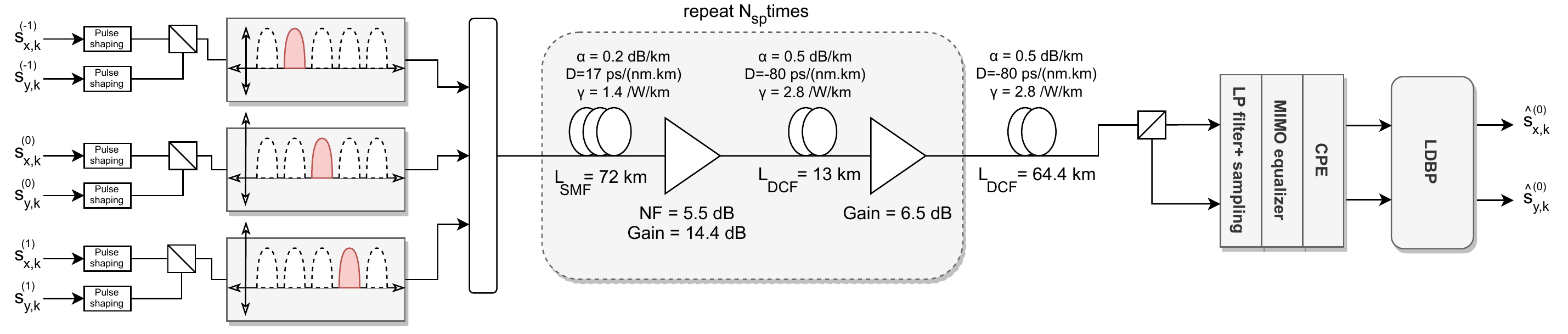}
    \caption{The dispersion-managed WDM optical fiber system with linear DSP and NN equalizer.}
    \label{fig:sysmodel}
\end{figure*}
In this paper, we study nonlinearity mitigation in legacy DM systems since that is still a relevant use case.
We consider the transmission of wavelength-division multiplexed (WDM) QAM signals with coherent-detection, copropagating with OOK signals in adjacent frequency bands.
First, we propose a variant of DBP with fractional number of steps/span that is adapted to DM systems. 
Next, we consider an LDBP where the parameters of the filters in the DBP are optimized. This yields a feed-forward NN that can be trained using gradient-based methods.
Lastly, we evaluate our proposed DBP and LDBP
by simulating a realistic dual-polarization WDM DM transmission
system taking into account losses, residual chromatic dispersion, Kerr nonlinearity, polarization-mode dispersion (PMD) and amplified spontaneous emission (ASE) noise. Q-factor improvements of 1.8 dB and 1.2 dB are obtained using LDBP with respect to linear equalization and DBP with similar complexity, respectively.
We show that the chromatic dispersion management reduces the complexity of the receiver DSP, enabling
nonlinearity mitigation using LDBP with few spatial segments.
\section{Back-propagation adapted to DM systems}

The propagation of signals in the two polarizations of optical fiber is described by the coupled nonlinear Schr\"odinger equation (CNLSE)
\begin{IEEEeqnarray}{rCl}
    \frac{\partial U_H(z,t)}{\partial z} &= & -\frac{\alpha}{2} U_H
    -\beta_1\frac{\partial U_H}{\partial t} - \frac{j\beta_2}{2}\frac{\partial^2 U_H}{\partial t^2} 
\nonumber \\
&&+ j\gamma\; \Bigl(|U_H|^2+\frac{2}{3}|U_{V}|^2\Bigr)U_H.
                            \label{eq:nlse}
\end{IEEEeqnarray}
Here, $U_H$ and $U_V$ represent the complex envelope of the signal as a function of time $t$ and distance $z$ in polarization $H,V \in \{X,Y\}$, $H\neq V$, $\alpha$ is the loss parameter, $\beta_{1}$ is the first-order dispersion coefficients that depends on the polarization, $\beta_{2}$ is the chromatic dispersion (CD) coefficient, and $\gamma$ is the nonlinearity parameter.

In DBP, the SSFM is applied to \eqref{eq:nlse}, averaged over a rapidly-varying state-of-polarization, with negated parameters. Discretize $U_H$ and $U_V$ to $\mathbf{X}\in\mathbb C^N$ and $\mathbf{Y}\in\mathbb{C}^N$, respectively;
the discrete-time model for the $\mathbf X$ signal considered in DBP is
\begin{equation} 
    \frac{d\mathbf{X}(z)}{dz} = \boldsymbol{B}\mathbf{X}(z)+\mathbf{K} (\mathbf X, \mathbf Y) \odot \mathbf{X}(z),
    \label{disc_NLSE}
\end{equation}
where $\odot$ is the Hadamard product, $\boldsymbol{B}=\boldsymbol{W}^{-1}\text{diag}(H_1,...,H_n)\boldsymbol{W}$, $\boldsymbol{W}$ is 
the discrete Fourier transform matrix, $H_k = \frac{\alpha}{2} - j\beta_1 \omega_k- j\beta_2 \omega_k^2/2$, 
$\omega_k=2\pi f_k$, $f_k$ is the $k$-th discrete frequency, and 
\begin{equation*}
\mathbf K(\mathbf X, \mathbf Y) = -j\frac{8}{9}\gamma \Bigl(\mathbf X \odot \mathbf{X}^*+\mathbf Y\odot\mathbf{Y}^*\Bigr).
\end{equation*}
The solution of the linear and nonlinear part of \eqref{disc_NLSE} in the spatial segment $[z, z+\delta]$ are 
$\mathbf{X}(z+\delta) = e^{\delta\mathbf{B}}\mathbf{X}(z)$ and $\mathbf{X}(z+\delta) = e^{\delta\mathbf{K}}\mathbf{X}(z) $, respectively. The equations for $\mathbf Y$ are similar.
\begin{figure}[!b]
    \centering
    \includegraphics[width = 0.93\linewidth]{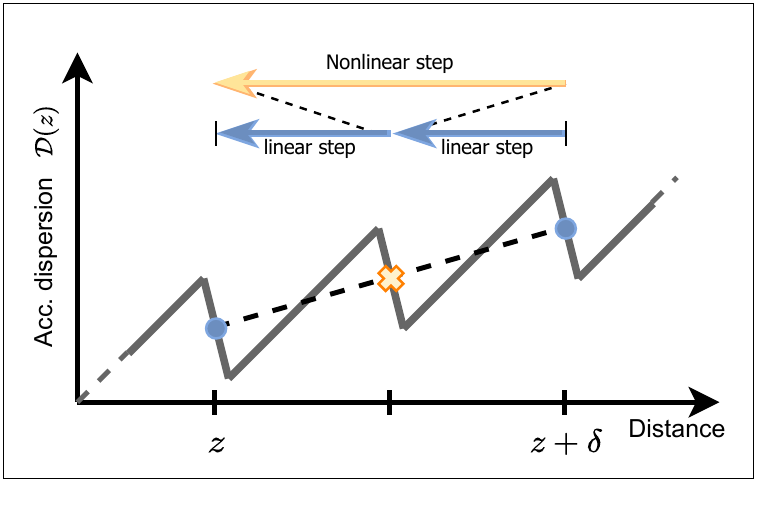}
    \caption{A single DBP step applied over two spans for a given dispersion map.}
    \label{fig:dispersionmap}
\end{figure}

The accumulated chromatic dispersion grows linearly with distance in uncompensated links. 
In DM systems, on the other hand, the dispersion introduced by the transmission fiber e.g., standard single mode fiber (SMF), is partially canceled at the end of each span via the application of DCF. 
The cascade of SMF and DCF forms a dispersion map with an accumulated dispersion $\mathcal{D}(z)$ as a function of distance $z$; see Fig.~\ref{fig:dispersionmap}. At the end of the link, usually a residual dispersion is left. Since DCF partially compensates for the accumulated dispersion, 
the linear step in the equalizer at the receiver has to account for the DCFs installed in the link, such that the total amount of compensated CD is equivalent to the total amount of dispersion generated by DCF and SMF together.
Consider a DBP step over segment $[z, z+\delta]$, and a dispersion map $\mathcal{D}(z)$ tracing the residual dispersion over the above segment, the linear step in the proposed DBP is adjusted such that
$\delta \mathbf B=\delta_1 \mathbf B_1+\delta_2 \mathbf B_2$,
where $\delta_1$ and $\delta_2$ are, respectively, the length of SMF and DCF, $\delta = \delta_1 + \delta_2$, and $\mathbf{B}_{\text{1}}$ and $\mathbf{B}_{\text{2}}$ are the time-domain update matrices with the coefficients of SMF and DCF.
The nonlinear step is similar to that used in unmanaged systems, with nonlinear coefficient $\gamma_{\text {eff}}$ averaged across the two types of fiber. It can be seen that this version of DBP does not restrict one to integer number of steps/span, which is important in real-time implementation of DBP in high-speed transmission.

\section{Learned digital back-propagation}

LDBP uses the SSFM channel model as a blue print for the complex-valued NN structure.
Denote the output of $m$-th layer by the time-domain vectors $\mathbf X^{(m)}$ and $\mathbf Y^{(m)}$, where the upper index $m$ refers to the step number.
The LDBP model takes two inputs $\mathbf X^{(0)}$ and $\mathbf Y^{(0)}$, and outputs $\mathbf X^{(M)}$ and $\mathbf Y^{(M)}$, corresponding to the equalized complex-value signals sampled at the  sampling rate of the input. Each layer applies a function $\mathbf X^{(m+1)} = \Phi(\mathbf X^{(m)}\circledast \mathbf A_m)$ to the input vector, where $\circledast$ denotes circular convolution, $\mathbf A_m$ is the weight vector of the $m$-th convolutional layer, $0\leq m \leq M$, and $\Phi(\mathbf X)$ is special activation function, we call ``\emph{Kerr activation}'', which performs the same nonlinear phase shift in DBP
\begin{equation*}
\Phi(\mathbf{X}) =~ \mathbf{X} \odot \exp\Bigl\{-j\frac{8}{9}\bar\gamma\delta \bigl(\mathbf X \odot \mathbf{X}^*+\mathbf Y\odot\mathbf{Y}^*\bigr)\Bigr\},
\end{equation*}
where $\bar\gamma$ is a trainable parameter, and $\exp(\cdot)$ is applied component-wise.
The transformation  is identical for the $y$-polarization, upon exchanging $\mathbf X$ and $\mathbf Y$.
The weights of the convolutional layers are initialized with the coefficients of the DBP linear step.

\section{Performance results}

The WDM-DP system and the LDBP NN are simulated using Python's TensorFlow library. WDM-DP-16QAM signals were generated at $32$~Gbaud using root raised cosine pulse-shape with $0.06$ roll-off factor. The optical fiber link consists of $N_{sp}=$~$28$ spans, each span including an SMF and a DCF measuring $72$~km and $13$~km, respectively, as indicated in Fig.~\ref{fig:sysmodel}. The length of DCF is chosen such that it compensates for $85$\% of the dispersion in each span. The dispersion map parameters are indicated in Table~\ref{fig:table}. The fiber is terminated with DCF, such that the total residual dispersion at RX is zero.
An amplification gain $G_{\text{SMF}} = 14.4$ dB is applied at the end of the SMF, followed by a gain of $G_{\text{DCF}} = 6.5$ dB after the DCF. The SMF parameters are: $\alpha$ = $0.2$~dB/km, $D=17$ ps/(nm-km), $\gamma$ = $1.4$ W$^{-1}$km$^{-1}$. DCF parameters are: $\alpha$ = $0.5$~dB/km, $D=-80$  ps/(nm-km), $\gamma$=~$2.8$ W$^{-1}$km$^{-1}$.

The central WDM channel is surrounded by $4$ channels with the frequency spacing $37.5$~GHz.
Signal propagation is simulated with $72$ steps/span for SMF, and $13$ steps/span for DCF.
The WDM signals are sampled at the high sampling rate of 16 samples/symbol, for accurate simulation. 
At the receiver, first a root raised-cosine band-pass filter is applied to filter out the adjacent channels, such that only the central channel is processed by DBP or LDBP. As a result, the channel of interest is affected by the interference introduced by the adjacent channels primarily via the cross-phase modulation. Next, the signal is down-sampled to twice the symbol rate before being processed by the conventional DSP chain which equalizes linear effects, such as CD, PMD and polarization mixing. DBP and LDBP equalizers are placed after the linear equalizer.

\begin{figure}[t]
\includegraphics[width=\linewidth]{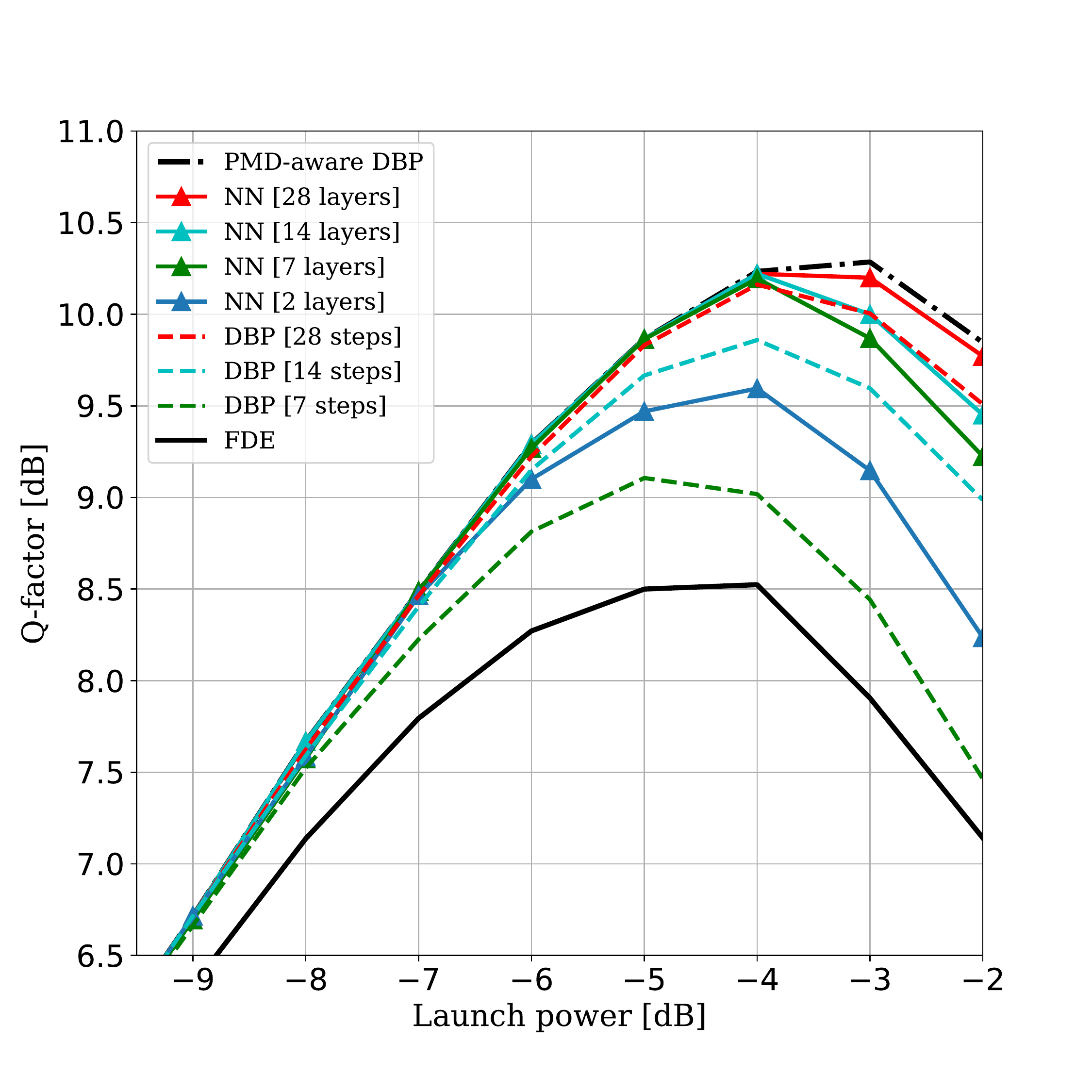}
\centering
\caption{Comparison between the achieved $Q$-factors for different equalization methods.}
\label{fig:Performance}
\end{figure}

\begin{figure}
     \centering
     \begin{subfigure}[b]{0.225\textwidth}
         \centering
         \includegraphics[width=\textwidth]{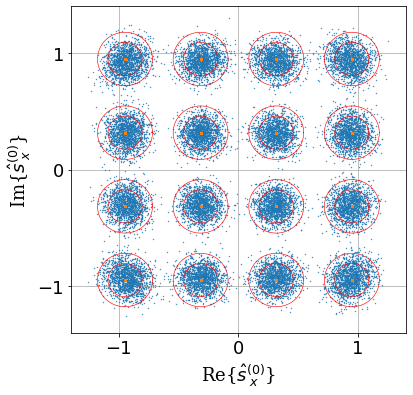}
         \caption{LDBP equalizer}
         \label{fig:LDBP}
     \end{subfigure}
     \begin{subfigure}[b]{0.225\textwidth}
         \centering
         \includegraphics[width=\textwidth]{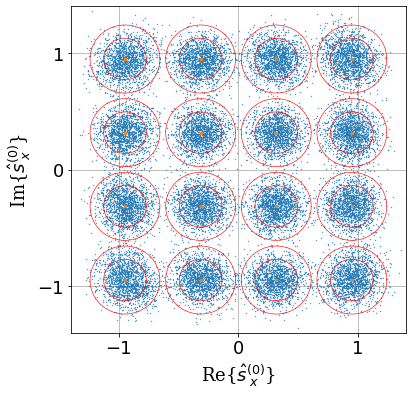}
         \caption{Linear Equalizer.}
         \label{fig:Lin}
     \end{subfigure}
        \vspace{10pt}
        \caption{Constellations at RX. The inner and outer circles respectively represent areas where 68\% and 95\% of symbols are detected.\vspace{6pt}}
        \label{fig:const}
\end{figure}

For training the NN, we generated $2^{15}$ input output signals at each launch power, with $2$ samples/symbol. The length of the input and output vectors is $512$ and $384$, respectively, to account for the channel memory. All layers were initialized with the corresponding parameters in a linear step of DBP at the full step size $\delta$, except for the first and last layers that correspond to a half  step $\delta/2$.

The NN performance, measured by the $Q$-factor obtained from the bit error rate, is shown in Fig.~\ref{fig:Performance}. 
The PMD-aware DBP curve is obtained by back-propagating the received signal under the assumption that the PMD coefficients are known to the receiver. This receiver is impractical, and is evaluated here merely to demonstrate an upper bound on the $Q$-factor of LDBP. 
Both DBP and LDBP are simulated with $M \in \{7,14,28 \}$ steps or layers.
LDBP with $7$ layers achieves the peak $Q$-factor = $10.2$ dB at launch power $P$=$-4$ dBm. Compared to the DBP with the same number of steps and launch power, this is a $1.2$~dB gain in $Q$-factor. The peak value for LDBP occurs at $1$ dB higher launch power. Moreover, LDBP with only $2$ layers achieved $9.5$ dB  peak $Q$-factor, $1$ dB higher than that of the linear equalizer at the same launch power.
Lastly, Figs.~\ref{fig:const} show the constellations at the output of the linear equalizer and at the output of the $28$-layer LDBP equalizer, at the peak $Q$-factor.
Overall, the results suggest that LDBP achieves a considerably higher $Q$-factor given a fixed number of spatial segments.

\begin{table}
 \centering
\begin{tabular}{|c|c|}
    \hline
    Dispersion pre-compensation & -1224 ps/nm \\
    \hline
    Residual dispersion/span & -184 ps/nm \\
    \hline
    Residual dispersion at RX  & 0 ps/nm\\
    \hline
  \end{tabular}
\vspace{0pt}
\caption{Dispersion map parameters.}
\label{fig:table}
\end{table}

\section{Conclusions}

We proposed a variant of DBP adapted for DM systems for the mitigation of nonlinear effects. We optimized the DBP parameters using an LDBP approach. The resulting LDBP outperforms linear equalization and DBP with the same computational complexity, by 1.8 and 1.2 in $Q$-factor.

\section{Acknowledgement}
This project has received funding from the European Union Horizon 2020 research and innovation programme, under the Marie Sk\l{}odowska-Curie grant agreement No. 813144.

\printbibliography

@ARTICLE{DSP_savory,
  author={Savory, Seb J.},
  journal={IEEE Journal of Selected Topics in Quantum Electronics}, 
  title={Digital Coherent Optical Receivers: Algorithms and Subsystems}, 
  year={2010},
  volume={16},
  number={5},
  pages={1164-1179},
  doi={10.1109/JSTQE.2010.2044751}
  }

@ARTICLE{Ezra_Ip_DBP, 
author={Ip, Ezra and Kahn, Joseph M.}, 
journal={Journal of Lightwave Technology},
title={Compensation of Dispersion and Nonlinear Impairments Using Digital Backpropagation},   
year={2008}, 
volume={26},
number={20}, 
pages={3416-3425},
doi={10.1109/JLT.2008.927791}}

@article{Opt_express_LDBP,
author = {Takashi Inoue and Ryosuke Matsumoto and Shu Namiki},
journal = {Opt. Express},
number = {9},
pages = {14851--14872},
publisher = {OSA},
title = {Learning-based digital back propagation to compensate for fiber nonlinearity considering self-phase and cross-phase modulation for wavelength-division multiplexed systems},
volume = {30},
month = {4},
year = {2022},
doi = {10.1364/OE.454841},}

@ARTICLE{Hager_butler_LDBP, 
author={B\"atler, Rick M. and H\"ager, Christian and Pfister, Henry D. and Liga, Gabriele and Alvarado, Alex}, 
journal={Journal of Lightwave Technology}, 
title={Model-Based Machine Learning for Joint Digital Backpropagation and PMD Compensation},  
year={2021}, 
volume={39}, 
number={4},
pages={949-959}, 
doi={10.1109/JLT.2020.3034047}}

@ARTICLE{Hager_PBDL, 
author={H\"ager, Christian and Pfister, Henry D.}, 
journal={IEEE Journal on Selected Areas in Communications},  
title={Physics-Based Deep Learning for Fiber-Optic Communication Systems},   year={2021},
volume={39}, 
number={1}, 
pages={280-294},  
doi={10.1109/JSAC.2020.3036950}
}

@ARTICLE{APTL_NN, 
author={Fan, Qirui and Lu, Chao and Lau, Alan Pak Tao}, 
journal={Journal of Lightwave Technology}, 
title={Combined Neural Network and Adaptive DSP Training for Long-Haul Optical Communications}, 
year={2021},  volume={39}, 
number={22}, 
pages={7083-7091},
doi={10.1109/JLT.2021.3111437}
}

@ARTICLE{Antonio_DBP,  
author={Napoli, Antonio and Maalej, Zied and Sleiffer, Vincent A. J. M. and Kuschnerov, Maxim and Rafique, Danish and Timmers, Erik and Spinnler, Bernhard and Rahman, Talha and Coelho, Leonardo Didier and Hanik, Norbert},
journal={Journal of Lightwave Technology},  
title={Reduced Complexity Digital Back-Propagation Methods for Optical Communication Systems},  
year={2014},
volume={32}, 
number={7},
pages={1351-1362}, 
doi={10.1109/JLT.2014.2301492}
}

@INPROCEEDINGS{Hagers_first,
author={H\"ager, Christian and Pfister, Henry D.},  
booktitle={2018 Optical Fiber Communications Conference and Exposition (OFC)}, 
title={Nonlinear Interference Mitigation via Deep Neural Networks},   
year={2018},
volume={}, 
number={}, 
pages={1-3}, 
doi={}
}

\vspace{-4mm}
-------------%
\end{document}